\documentclass[pre,aps,twocolumn,amsmath,amssymb]{revtex4}

\pdfoutput=1
\usepackage{hyperref}
\usepackage{graphicx}
\usepackage[usenames]{color}

\def\cal#1{\mathcal{#1}}
\def\eqq#1{Eq.~(\ref{#1})}

\def\f#1{Fig.~\ref{#1}}

\def\c#1{~\cite{#1}}
\def\cc#1{Ref.\c{#1}}

\newcommand{\con}[1]{\textcolor{red}{#1}}

\def\beq{\begin{equation}}
\def\eeq{\end{equation}}
\def\bea{\begin{eqnarray}}
\def\eea{\end{eqnarray}}

\def\kt{k_{\rm B}T}

\begin{document}

\title{Hierarchical assembly may be a way to make large information-rich structures}

\author{Stephen Whitelam\footnote{\texttt{swhitelam@lbl.gov}}}
\affiliation{Molecular Foundry, Lawrence Berkeley National Laboratory, 1 Cyclotron Road, Berkeley, CA 94720, USA}
\begin{abstract}

Self-assembly in the laboratory can now yield `information-rich' nanostructures in which each component is of a distinct type and has a defined spatial position. Ensuring the thermodynamic stability of such structures requires inter-component interaction energies to increase logarithmically with structure size, in order to counter the entropy gained upon mixing component types in solution. However, self-assembly in the presence of strong interactions results in general in kinetic trapping, so suggesting a limit to the size of an (equilibrium) structure that can be self-assembled from distinguishable components. Here we study numerically a two-dimensional hierarchical assembly scheme already considered in experiment. We show that this scheme is immune to the kinetic traps associated with strong `native' interactions (interactions designed to stabilize the intended structure), and so, in principle, offers a way to make large information-rich structures. In this scheme the size of an assembled structure scales exponentially with the stage of assembly, and assembly can continue as long as random motion is able to bring structures into contact. The resulting superstructure could provide a template for building in the third dimension. The chief drawback of this scheme is that it is particularly susceptible to kinetic traps that result from `non-native' interactions (interactions not required to stabilize the intended structure); the scale on which such a scheme can be realized therefore depends upon how effectively this latter kind of interaction can be suppressed.
\end{abstract}
\maketitle

\section{Introduction}
\label{intro}

Molecular self-assembly in the laboratory is a promising way of making useful materials\c{glotzer2007anisotropy,barth2005engineering,whitesides2002self,blake1999inorganic}. Self-assembly mediated by DNA\c{seeman1998dna}, in particular, has been used to create a wide range of nanostructures\c{winfree1998design,wei2012complex,fujibayashi2007toward}, some of which can perform basic functions~\cite{andersen2009self}. Recent work has demonstrated the self-assembly of DNA `brick' nanostructures, which are solid, equilibrium structures in which each component or brick is of a distinct type and has a defined position\c{ke2012three,ke2014dna}. These structures self-assemble in solution because inter-component interactions, which are mediated by DNA basepairs, are stronger between components designed to be adjacent in the assembled structure than between components not designed to be adjacent in the assembled structure. A growing body of theoretical work\c{fujibayashi2009error,schulman2004one,chen2005error,winfree2004proofreading,rothemund2004algorithmic,halverson2013dna,reinhardt2014numerical,hedges2014growth,murugan2015undesired,jacobs2015communication} indicates that this principle of component-type complementarity might be used quite generally to create defined, multicomponent assemblies of e.g. colloids or other nanoscale building blocks. `Information-rich' materials\c{zuckermann2011peptoid} of this kind have considerable technological potential\c{frenkel2015order,cademartiri2015programmable}.

\begin{figure*}[t]
\centering
\includegraphics[width=\linewidth]{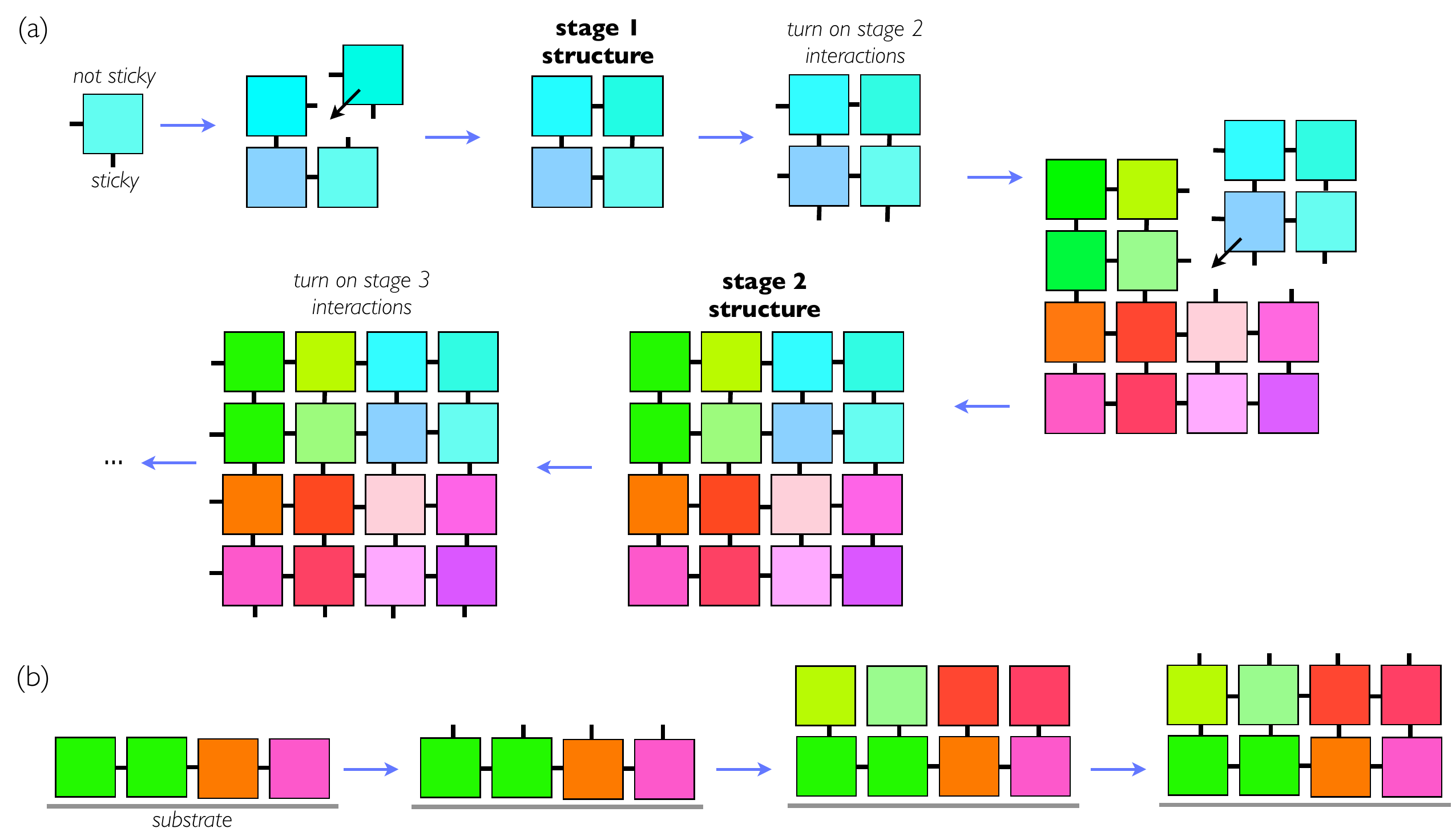}
\caption{\label{fig1} Self-assembly of large, information-rich structures in thermal equilibrium will likely require strong inter-component interactions\c{hedges2014growth,murugan2015undesired,frenkel2015order}. (a) The hierarchical assembly scheme demonstrated experimentally in \cc{park2006finite} is immune to the kinetic traps associated with strong native interactions, i.e. interactions designed to stabilize the assembled structure. Each stage of assembly involves the formation of squares, mediated by four chemically-specific internal bonds. No matter the strength of these bonds or the order in which they form, kinetic trapping in the form of mis-binding cannot occur. (b) The resulting two-dimensional assembly could provide a template for building a three-dimensional structure, which could in principle also be done in a way that avoids kinetic traps associated with strong native interactions.}
\end{figure*}

However, theoretical work also suggests that some obstacles must be overcome before we can self-assemble information-rich structures of arbitrary size (DNA brick nanostructures are about 1000 components in size\c{ke2012three,ke2014dna}). For one, the growth rate of a structure made of $Q$ distinguishable components will likely be of order $Q$ times less than that of the corresponding indistinguishable-particle structure, because only about 1 in $Q$ inter-component collisions can promote growth\c{hedges2014growth}. For another, the energy scale of inter-component bonds must grow on the scale of $\kt \ln Q$ in order to render the desired structure thermodynamically stable, because bond energies must counter the entropy associated with permuting component types in solution\c{hedges2014growth,murugan2015undesired,frenkel2015order} (see Appendix A). This energy scale is, for macroscopic objects with $Q \sim 10^{24}$, in excess of $50\, \kt$. Molecular self-assembly in the presence of interaction energies large on the scale of $\kt$ generally results in kinetic trapping, because strong bonds can prevent the correction of mistakes that happen when components undergoing Brownian motion collide randomly\c{jack2007fluctuation,hagan2006dynamic,new_rapaport,wilber2007reversible,zaccarelli2007colloidal,lu2008gelation,rob}. This is so even if the only interactions present are those designed to stabilize the intended structure -- we will call these {\em native} interactions -- so suggesting a limit to the scale on which information-rich structures may be self-assembled.

To overcome this limitation one must either seek to make defined {\em nonequilibrium} structures\c{sanz2007evidence,peters2009competing,scarlett2010computational,whitelam2014self}, or else engineer an assembly pathway that is immune to the kinetic traps associated with strong native inter-component interactions. This paper focuses on the second of these options. 

We will focus on a 2D assembly scheme, a version of which has been used in experiment to create DNA nanotile lattices\c{park2006finite} (that scheme aimed to produce many copies of a small target structure; here we will be concerned with making one copy of a large target structure). In this scheme, each stage of assembly involves the formation of distinct squares, stabilized by four internal bonds. Each bond is in general mediated by multiple components. In the presence of native interactions only, each square can be formed only from four particular pieces. No matter the strength of the bonds between these four pieces, or the order in which bonds are made, no mis-binding can happen during construction of the square. In other words, this scheme is immune to the kinetic traps associated with strong native inter-component interactions. To enable such an assembly pathway one can either combine reactants in stages\c{park2006finite}, or `switch on' inter-component interactions in stages, as illustrated in \f{fig1}(a). In the first stage of assembly, selected interactions between monomers promote the assembly of squares of size 4. One then turns on additional interactions to promote the formation of second-stage structures, squares of size 16, and so on, with stage $n$ structures being of size $2^{2n}$. This process can continue as long as structures can be brought into contact, so permitting the generation of an arbitrarily large information-rich array. This array could provide a template for building in the third dimension, with e.g. subsequent layers deposited one at a time\c{cademartiri2015programmable}; see panel (b).

The chief drawback of this scheme it that it is, like other forms of hierarchical assembly\c{villar2009self,haxton2013hierarchical,madge2015design}, particularly susceptible to kinetic traps caused by `undesigned' or {\em non-native} interactions, by which we mean interactions that are not required to stabilize the intended structure. Such interactions include those between components {\em not} designed to be adjacent in the assembled structure, or interactions between intended neighbors that promote binding incompatible with the target structure. Non-native interactions might arise in experiment because of `accidental' complementarity between DNA sequences, or because of nonspecific van der Waals attractions. `Non-hierarchical' many-component self-assembly can proceed in the presence of such interactions\c{ke2012three,reinhardt2014numerical}, provided that the difference in energy scales between the set of native and non-native interactions is large enough\c{hedges2014growth,murugan2015undesired}. But attractive non-native interactions are tolerated poorly by a stage-by-stage hierarchical scheme, in which the effective building block size is ever-increasing, because undesirable interactions between two clusters increase in strength in proportion to their surface area. Therefore, above some size scale, kinetic trapping of a hierarchical assembly scheme in the presence of attractive non-native interactions would seem to be inevitable. 
\begin{figure*}[t]
\centering
\includegraphics[width=\linewidth]{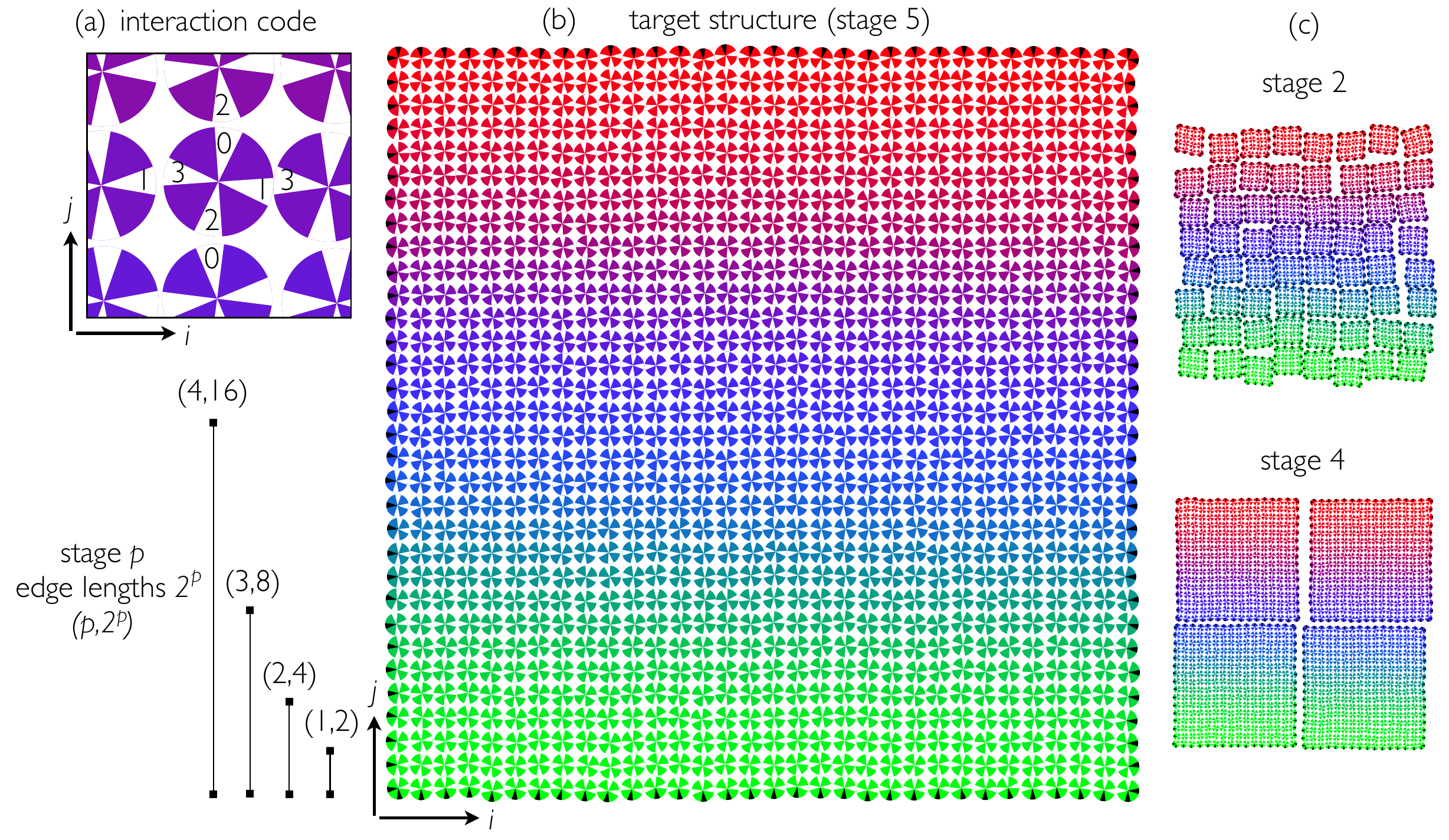}
\caption{\label{fig2} Simulation model of the idea sketched in \f{fig1}. (a) Disc geometry and patch-patch native interaction rules. (b) Target structure for self-assembly, a `stage 5' square of size $2^{10}=1024$. All discs in the structure are of a distinct type, as indicated by their color. (c) Maximally-bonded configurations of stage 2 and stage 4. }
\end{figure*}

In view of these considerations, the hierarchical scheme is likely to be inferior to a non-hierarchical scheme if non-native interactions are not intentionally suppressed. Indeed, the authors of \cc{park2006finite} found for DNA tiles that the `square-upon-square' hierarchical assembly scheme was inferior to a non-hierarchical scheme (in which all component types were combined directly), even when used to make a relatively small $4 \times 4$ array. Despite this finding, we argue here that the hierarchical scheme can in principle be used to build larger information-rich structures than can the non-hierarchical scheme (which cannot tolerate strong native interactions), {\em if} one can suppress non-native interactions. Perhaps this could be done by combining DNA-mediated interactions with some form of inter-particle repulsion (we discuss this point further in the conclusions section). If so, the stage-by-stage hierarchical assembly pathway offers a potential route to the assembly of large information-rich structures in thermal equilibrium.

In Section~\ref{model} we introduce an off-lattice particle-based computer model that can be used to study stage-by-stage hierarchical assembly. In Section~\ref{results} we show that simulations of this model confirm that stage-by-stage assembly results in error-free formation of a desired structure, even in the presence of arbitrarily strong native interactions. We show that the same scheme fares poorly in the presence of attractive non-native interactions. We also discuss the efficiency of a conventional self-assembly process in the presence of irreversible or reversible native interactions. We conclude in Section~\ref{conc}.

\section{Modeling stage-by-stage assembly}
\label{model}
\begin{figure*}[t]
\centering
\includegraphics[width=\linewidth]{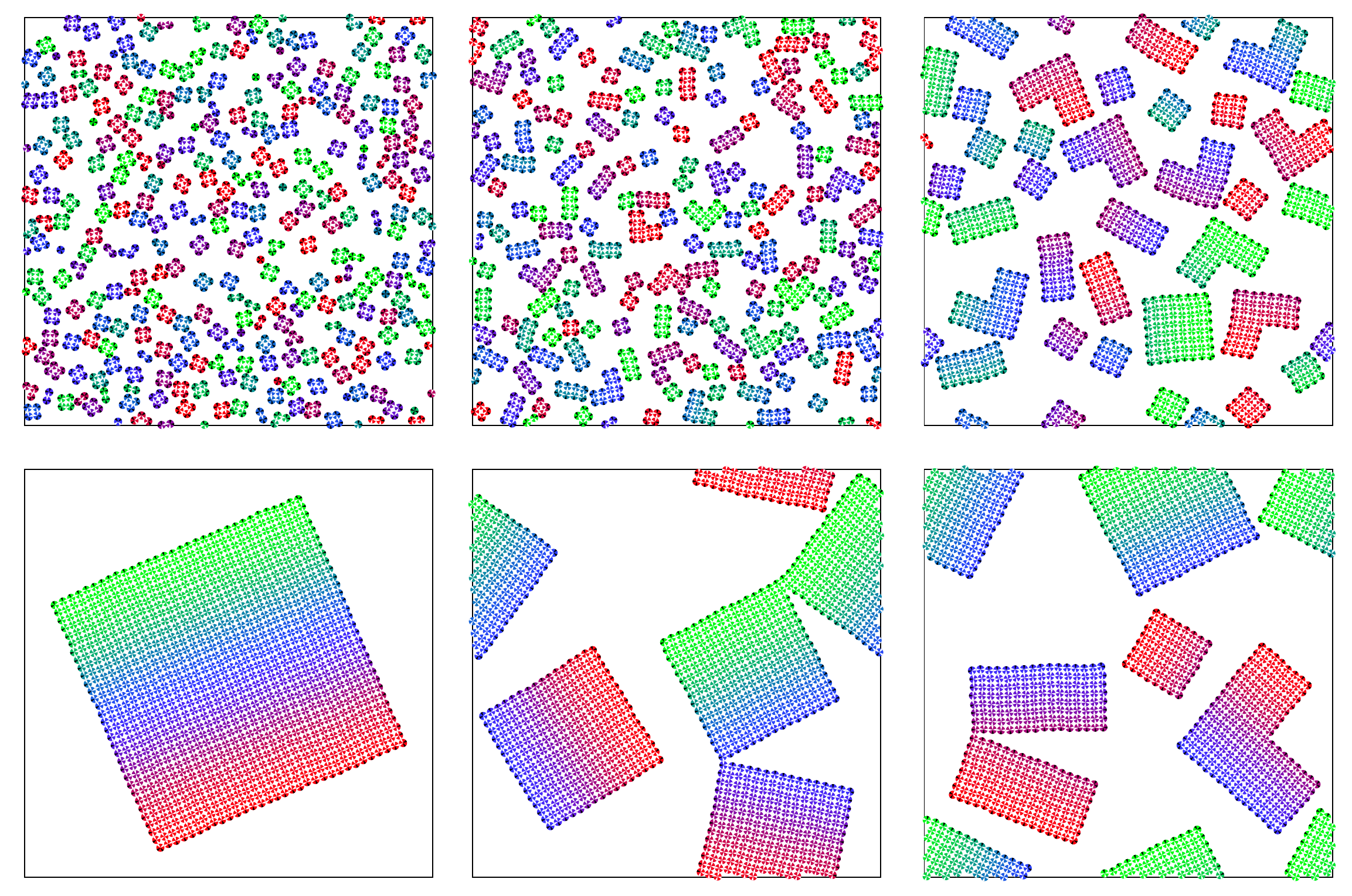}
\caption{\label{fig_hierarchical} A time-ordered series of snapshots from a single trajectory of the stage-by-stage hierarchical assembly procedure described in Section~\ref{model}, modeled on the experiments reported in \cc{park2006finite}. The procedure results in error-free assembly of the target structure (see \f{fig2}). Times of snapshots, clockwise from top left, are (in millions of MC sweeps) 5, 15, 50, 117, 163, 332.}
\end{figure*}

\begin{figure}[]
\centering
\includegraphics[width=\linewidth]{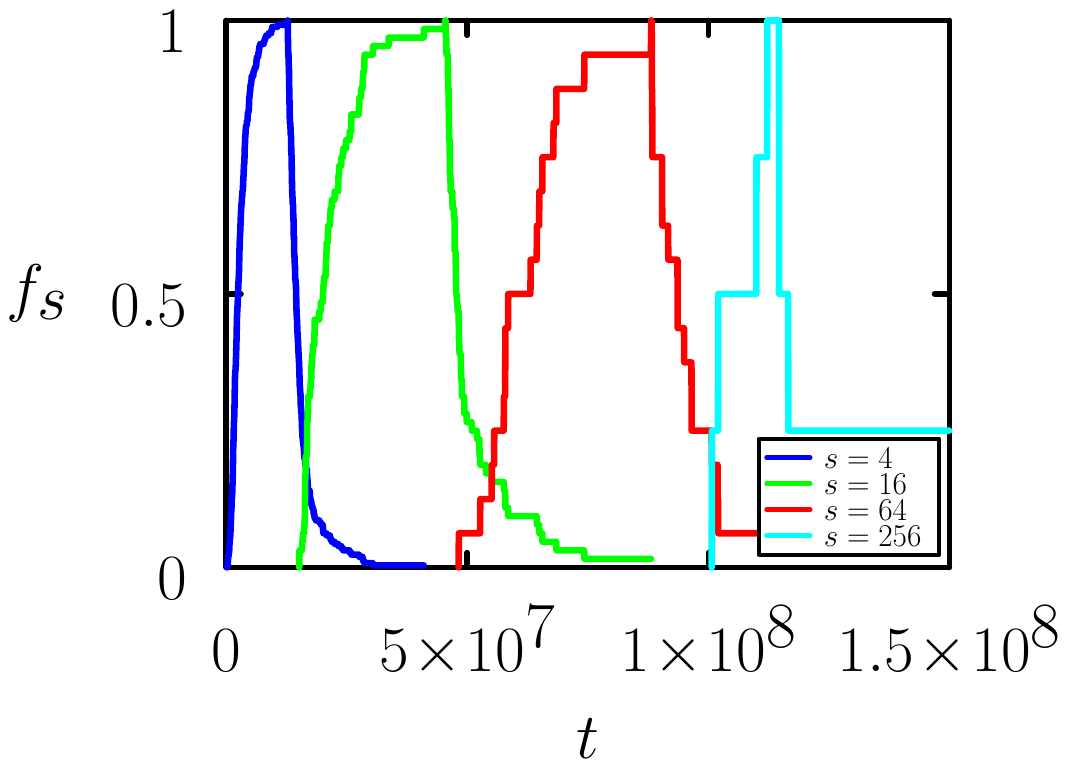}
\caption{\label{fig_hierarchical_graph} For the trajectory shown in \f{fig_hierarchical}, we plot as a function of time $t$ the fraction $f_{\rm s}$ of the system's mass contained within (natively-bonded) clusters of size $s$. The completion of each stage of assembly corresponds to the point at which each line reaches the value unity, at which point the interactions required to promote the next stage of assembly are automatically switched on. Assembly went to completion at time $t \approx 3 \times 10^8$.}
\end{figure}

The idea of assembling structures in a stage-by-stage manner has been studied theoretically in the DNA tile literature\c{demaine2011one,patitz2010self,abel2010shape}, where the possibility of extending stage-by-stage assembly to large scales was noted.  The particular case of square-upon-square assembly sketched in \f{fig1}(a) has been demonstrated in experiments in which DNA tiles formed many distinct $4 \times 4$ arrays\c{park2006finite}. Here we study this process numerically, using off-lattice molecular simulation, in order to demonstrate that it can in principle be used to avoid kinetic traps associated with very strong native interactions. For the reasons discussed in Section~\ref{intro}, such avoidance is a necessary feature of a process designed to make large information-rich structures in thermal equilibrium. We shall also show that this scheme is particularly vulnerable to kinetic trapping caused by non-native interactions.

Our model of stage-by-stage assembly comprises 1024 hard discs of radius $a$ on a smooth, two-dimensional substrate. Each disc is of a distinct type, one of 1024 possibilities, labeled $(i,j)$, with $i,j = 0,1,\dots, 31$. Discs are decorated with 4 sticky patches, each of opening angle $\pi/6$, arranged in a regular way (i.e. neighboring patch bisectors are separated by an angle $\pi/2$), so as to allow self-assembly of a square lattice structure. As shown in \f{fig2}(a), patches $p$ are numbered 0, 1, 2, 3 in a clockwise fashion. Discs possess attractive pairwise interactions that are square-well in both angle\c{kern2003fluid} and range. If the centers of two discs $(i,j)$ and $(i',j')$ are separated by a distance $d$ that satisfies $2a < d\leq 2a+a/5$, and if the line joining disc centers cuts through one patch on each disc, then discs receive a pairwise energetic reward of $-E \, \kt$, where
\begin{widetext}
\beq
E=f(i,j,p;i',j',p') \epsilon_{\rm native} +\left(1-f(i,j,p;i',j',p')\right)\epsilon_{\rm non-native}.
\eeq
\end{widetext}
Here $\epsilon_{\rm native}>0$ and $\epsilon_{\rm non-native}>0$ are the energy scales of native and non-native interactions. Unless otherwise stated we took $\epsilon_{\rm native} \to \infty$ and $\epsilon_{\rm non-native}=0$. The function $f$ depends on the disc- and patch identities involved in the pairwise contact: $p$ is the patch number of disc $(i,j)$ involved in the pairwise contact, and $p'$ is the patch number of disc $(i',j')$ involved in the pairwise contact. $f$ is chosen so that the structure shown in \f{fig2}(b) is the thermodynamically stable one. This structure, the target for self-assembly, is a square lattice in which disc type $(i,j)$ is found at the intersection of the the $i$th column and $j$th row of the lattice, numbered from the bottom left. That is, the bottom left disc is of type (0,0), and the top right disc is of type (31,31). Disc types are colored so that the target structure possesses a green-to-blue-to-red color gradient. To stabilize this structure we require $f(i,j,p;i',j',p')$ to be 1 if patch 1 on disc $(i,j)$ interacts with patch 3 on disc $(i+1,j)$, or if patch 0 on disc $(i,j)$ interacts with patch 2 on disc $(i,j+1)$; these are `native' interactions. Otherwise $f$ is 0, indicating a `non-native' interaction. 

In equation form these requirements read
\bea
\label{direct}
f(i,j,p;i',j',p') = \delta(i+1,i') \delta(j,j') \delta(p,1) \delta(p',3) \nonumber \\
+ \delta(j+1,j')\delta(i,i') \delta(p,0) \delta(p',2),
\eea
where $\delta(a,b)$ is 1 if $a=b$, and 0 otherwise. In pictures, in general, patches are shown white when engaged in a native contact, and black when unbound or engaged in a non-native contact.

To allow assembly of the target structure in a hierarchical manner we further distinguish disc native interactions by `stage'. Interactions up to stage $n$ allow the self-assembly of distinct squares of $2^{2n}$ discs. Thus, interactions of stage 1 can promote the formation of squares of 4 discs (of 256 possible types); stage 1 and 2 interactions allow the formation of 64 distinct square 16-mers; stage 1, 2, and 3 interactions allow the formation of 16 distinct square 64-mers; stage 1, 2, 3 and 4 interactions allow the formation of 4 distinct square 256-mers; and stage 1,2,3,4 and 5 interactions allow the formation of one square of size 1024. Maximally-bonded configurations that result from interactions of all stages up to 2, 4 and 5 are shown in \f{fig2}(b,c). 
\begin{figure*}[t]
\centering
\includegraphics[width=0.9\linewidth]{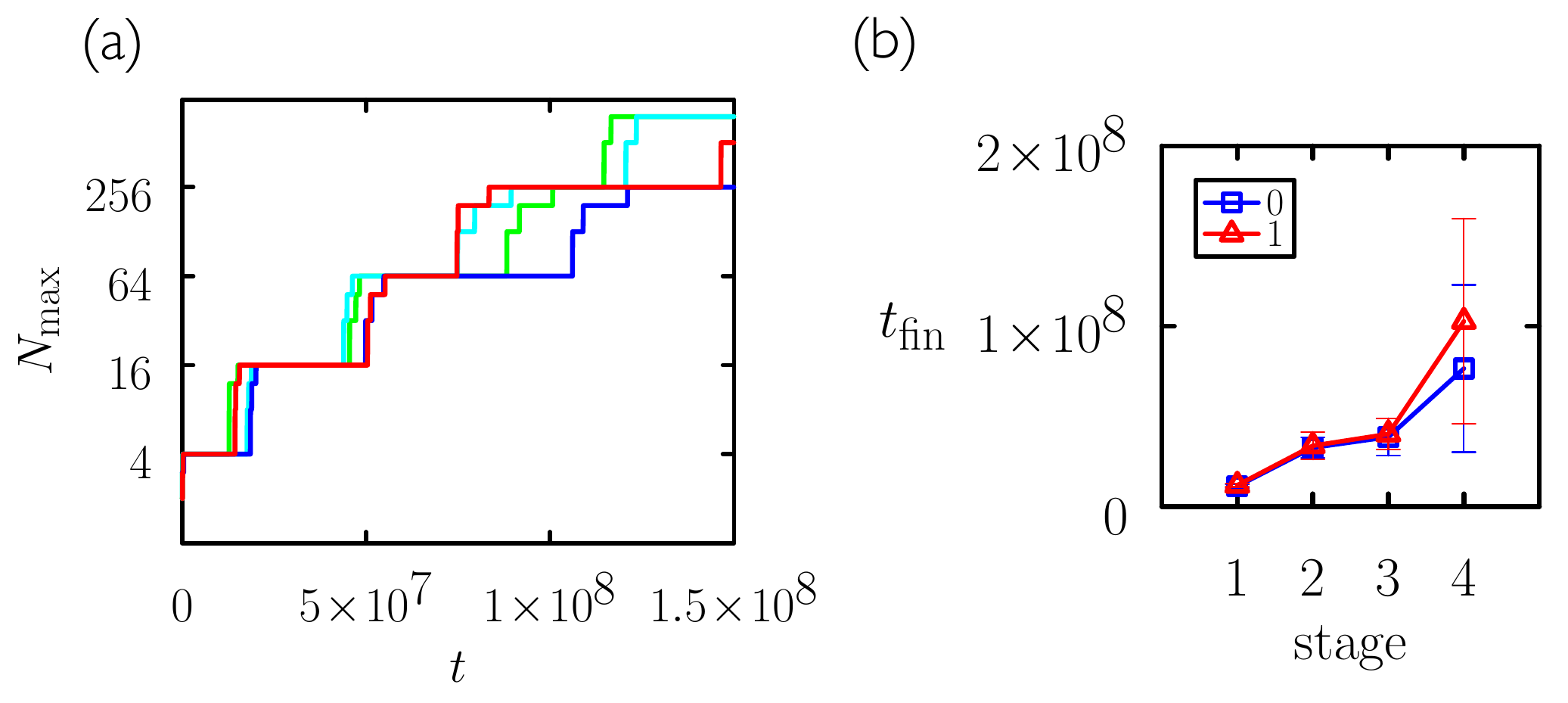}
\caption{\label{fig_hierarchical_graph_time} (a) Size $N_{\rm max}$ of the largest natively-bonded cluster, shown for 4 independent trajectories of the stage-by-stage hierarchical assembly scheme. (b) Mean (symbols) and standard deviation (error bars) of the time $t_{\rm fin}$ required to finish each stage of assembly (the times between the peaks of adjacent curves in a plot such as \f{fig_hierarchical_graph}), averaged over 8 independent trajectories. The blue and red lines correspond to cases in which non-native interaction strengths were set to zero and to $\kt$, respectively.}
\end{figure*}

In equation form we distinguish interactions that participate in different stages of assembly by modifying \eqq{direct} to read
\begin{widetext}
\beq
\label{hierarchical}
f(i,j,p;i',j',p') = \delta(i+1,i') \delta(j,j') \delta(p,1) \delta(p',3) \prod_{n=1}^5 \cal{M}_n(i') 
+ \delta(j+1,j')\delta(i,i') \delta(p,0) \delta(p',2) \prod_{n=1}^5 \cal{M}_n(j'),
\eeq
\end{widetext}
where $\cal{M}_n(i)=G_n$ if $i$ is a multiple of $2^{n-1}$, and is unity otherwise. The parameter $G_n$, which we control externally, is either 0, signaling that stage $n$ interactions are `turned off', or 1, signaling that stage $n$ interactions are `turned on'. We shall restrict ourselves to the case in which $G_n = 1$ implies $G_m = 1$ for all $m<n$, e.g. if stage 3 interactions are `turned on' ($G_3=1$), then so too are stage 1 and stage 2 interactions $(G_1=G_2=1)$. 

We did simulations of 1024 discs, one of each type, in the $NVT$ ensemble. Simulation boxes were square, with periodic boundaries, of size such that the disc hard-core packing fraction was 32\%. To demonstrate the fact that error-free hierarchical assembly can occur in the presence of strong interactions we took $\epsilon_{\rm native} \to \infty$, so that contacts, once made, could not be broken. Disc structures therefore do not dissociate, although they possess internal flexibility on account of the finite range and angular specificity of disc-disc interactions. Discs were initially randomly dispersed and oriented, subject to there being no hard-core overlaps. We used the virtual-move Monte Carlo algorithm\c{whitelam2007auk} described in the appendix of \cc{whitelam2009rcm} to move discs according to an approximation of overdamped motion. Translation and rotation of interacting clusters of discs occurs under this dynamical scheme, and the rate of cluster diffusion can be controlled to a degree by the user. We chose to reject moves of clusters of hydrodynamic radius $R$ so that, approximately, translational and rotational cluster diffusion constants scaled as $D_{\rm trans}(R) \propto R^{-1}$ and $D_{\rm rot}(R) \propto R^{-3}$ (see the SI of \c{haxton2012design}). Different choices may be appropriate for different types of surface. We chose the basic scale of disc displacement so that an isolated disc will diffuse a characteristic length equal to its own diameter in 33.3 Monte Carlo sweeps.

To promote stage-by-stage assembly we began simulations with only stage 1 interactions turned on, i.e. we set $G_1=1$ and $G_n =0$ for $n>1$. When the first stage of assembly was complete, i.e. when the simulation box contained 256 distinct square 4-mers, we turned on stage 2 interactions, i.e. we set $G_2=1$ (with the condition $G_1=1$ unchanged). Stage 2 interactions allow the 256 square 4-mers to assemble hierarchically into 64 distinct square 16-mers. When stage 2 of self-assembly was complete we turned on stage 3 interactions, and so on until stage 5 of self-assembly was complete. For comparison, we also carried out hierarchical simulations in which non-native patch-patch interactions (those that return zero on the right-hand side of \eqq{hierarchical}) were nonzero (\f{fig_hierarchical_graph_time}(b), \f{fig_undesigned_graph}, \f{fig_undesigned}), and we did `non-hierarchical' simulations in which all native interactions were turned on from the start of the simulation, i.e. we set $G_n =1$ for $n=1,2,3,4,5$ (\f{fig_sequential_strong}, \f{fig_non_hierarchical}, \f{fig_non_hierarchical_graph}).

\section{Simulation results}
\label{results}
\begin{figure*}[t]
\centering
\includegraphics[width=\linewidth]{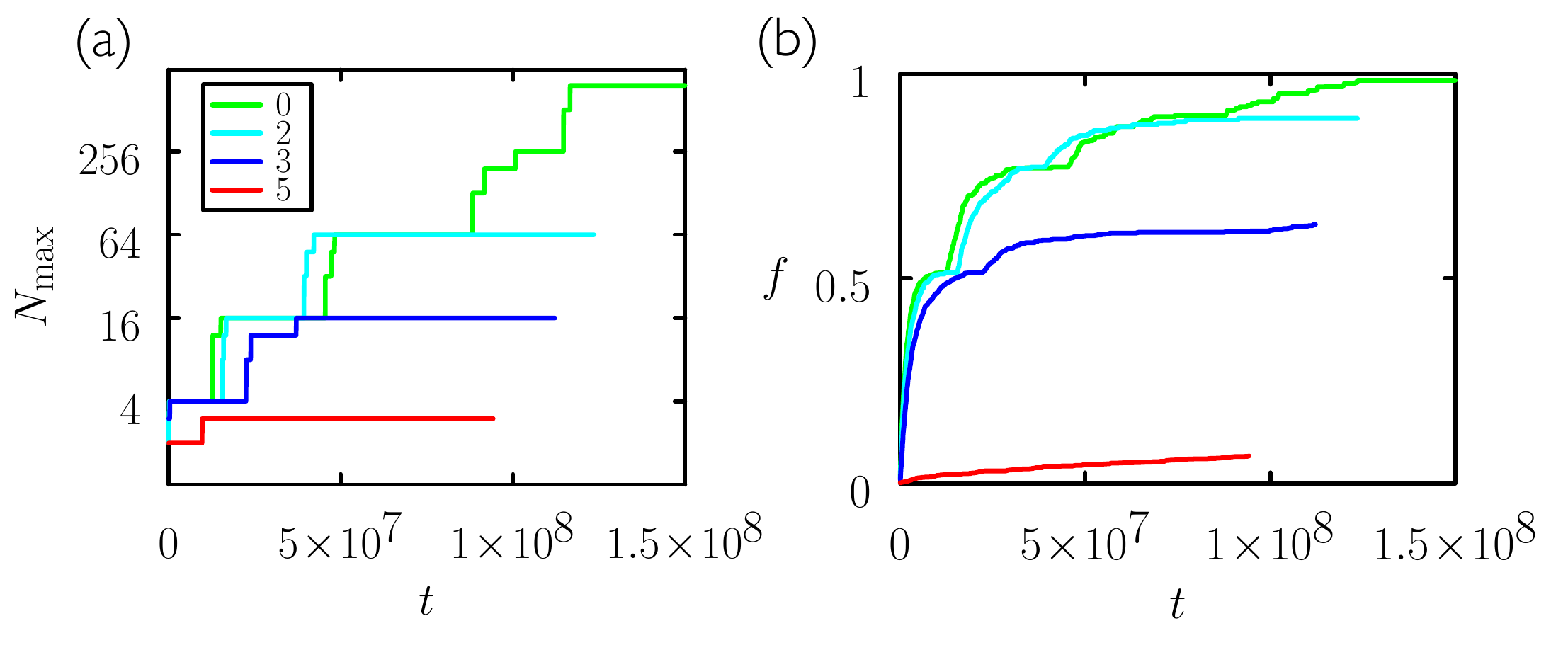}
\caption{\label{fig_undesigned_graph} (a) Size $N_{\rm max}$ of the largest natively-bonded cluster in the system from 4 trajectories run in the presence of attractive non-native interactions of $0\, \kt$, $2 \, \kt$, $3\,  \kt,$ and $5 \, \kt$. In all cases, native contacts are unbreakably strong. Assembly in the presence of non-native interactions becomes very slow at some point, the earlier the stronger the interaction. (b) Fraction $f \leq1$ of possible native contacts made at time $t$, for the same 4 trajectories.}
\end{figure*}

\begin{figure*}[t]
\centering
\includegraphics[width=\linewidth]{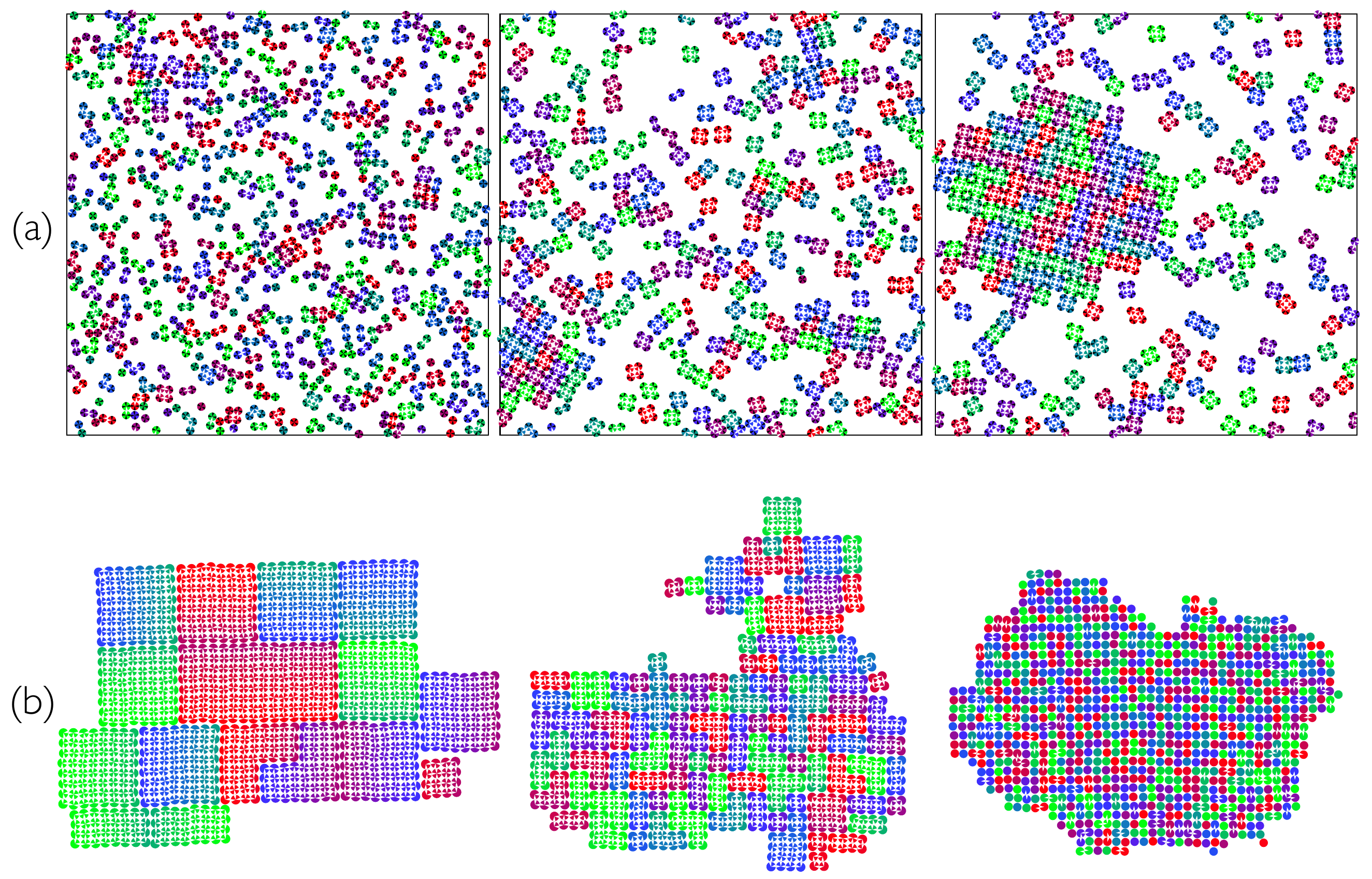}
\caption{\label{fig_undesigned} (a) A time-ordered series of snapshots (time increases left to right) from a simulation run in the presence of attractive non-native contacts of energy $3 \, \kt$. The square-upon-square hierarchical dynamics is overcome by a second organizational process that sees 4-mers assemble into compositionally-disordered structures. As a result, the hierarchical assembly procedure all but stops. (b) Structures assembled in the presence of attractive non-native contacts of energy $2 \, \kt$, $3 \, \kt$, and $5 \, \kt$ (the only patches shown are those engaged in native contacts).}
\end{figure*}

{\bf Stage-by-stage assembly works in the presence of arbitrarily strong native interactions.} In \f{fig_hierarchical} we show snapshots of simulation configurations, ordered clockwise from the top left, from a single dynamic trajectory of the stage-by-stage assembly procedure described in Section~\ref{model}. The scheme works -- the result is the low-energy, 1024-particle structure with no errors -- even though native interactions are unbreakably strong (non-native interactions are absent). Assembly is a diffusion-limited process, and no mis-bound configurations exist. This scheme is also unaffected by the `monomer starvation' kinetic trap seen in the study of viral capsid self-assembly\c{endres2002model,hagan2006dynamic} or in models of DNA brick assembly if one aims to produce multiple copies of a target structure\c{madge2015design}: all pieces have a prescribed set of binding partners, and so it is not possible to use up the monomer supply by making (say) a large number of trimers that cannot be completed. One needs to wait long enough for components to find their native partners, but the result is guaranteed to be free of error. 

In \f{fig_hierarchical_graph} we show, as a function of time (number of Monte Carlo sweeps), the fraction $f_{\rm s}$ of the system's mass contained within natively-bonded clusters of size $s$. The completion of each stage of assembly corresponds to the point at which the individual lines reach the value unity, at which point the interactions required to promote the next stage of assembly are automatically switched on. From trajectory to trajectory we found that that time required to complete each stage varied, the more so as assembly progressed. This variability can be seen in \f{fig_hierarchical_graph_time}(a), which shows for 4 independent trajectories the size of the largest (natively-bonded) cluster as a function of time. In panel (b) we show the mean and standard deviation of the time required to complete each stage of assembly, averaged over 8 independent trajectories. Stage completion time generally increases as assembly progresses, because larger clusters diffuse more slowly than smaller ones, even though the effective number of component types diminishes (see Appendix B).

\begin{figure}[t]
\centering
\includegraphics[width=\linewidth]{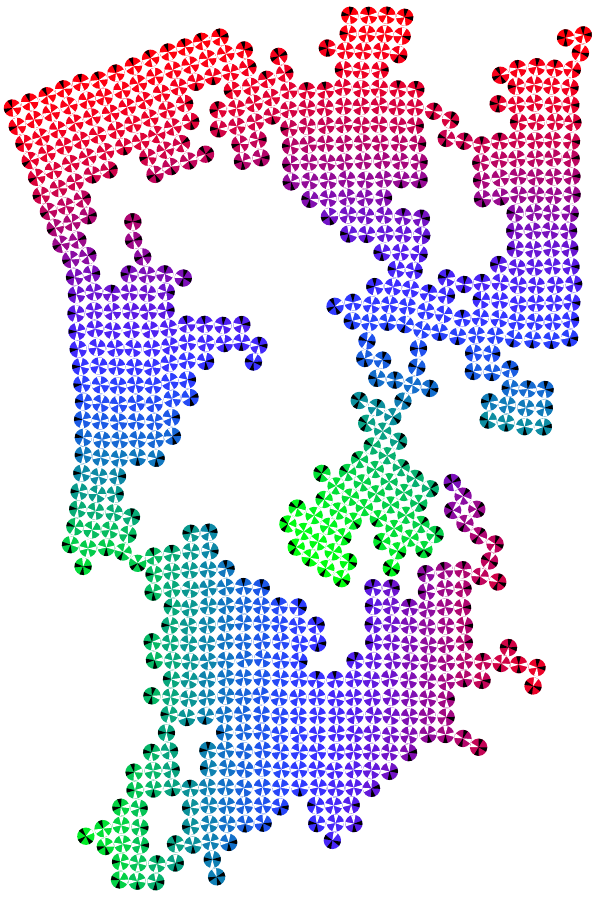}
\caption{\label{fig_sequential_strong} Non-stage-by-stage self-assembly results in kinetic trapping in the presence of strong native interactions. Here we show a structure that results from self-assembly in the presence of the same set of interactions use to produce \f{fig_hierarchical}; here, however, all stages of interaction were `switched on' from the start.}
\end{figure}

\begin{figure*}[t]
\centering
\includegraphics[width=\linewidth]{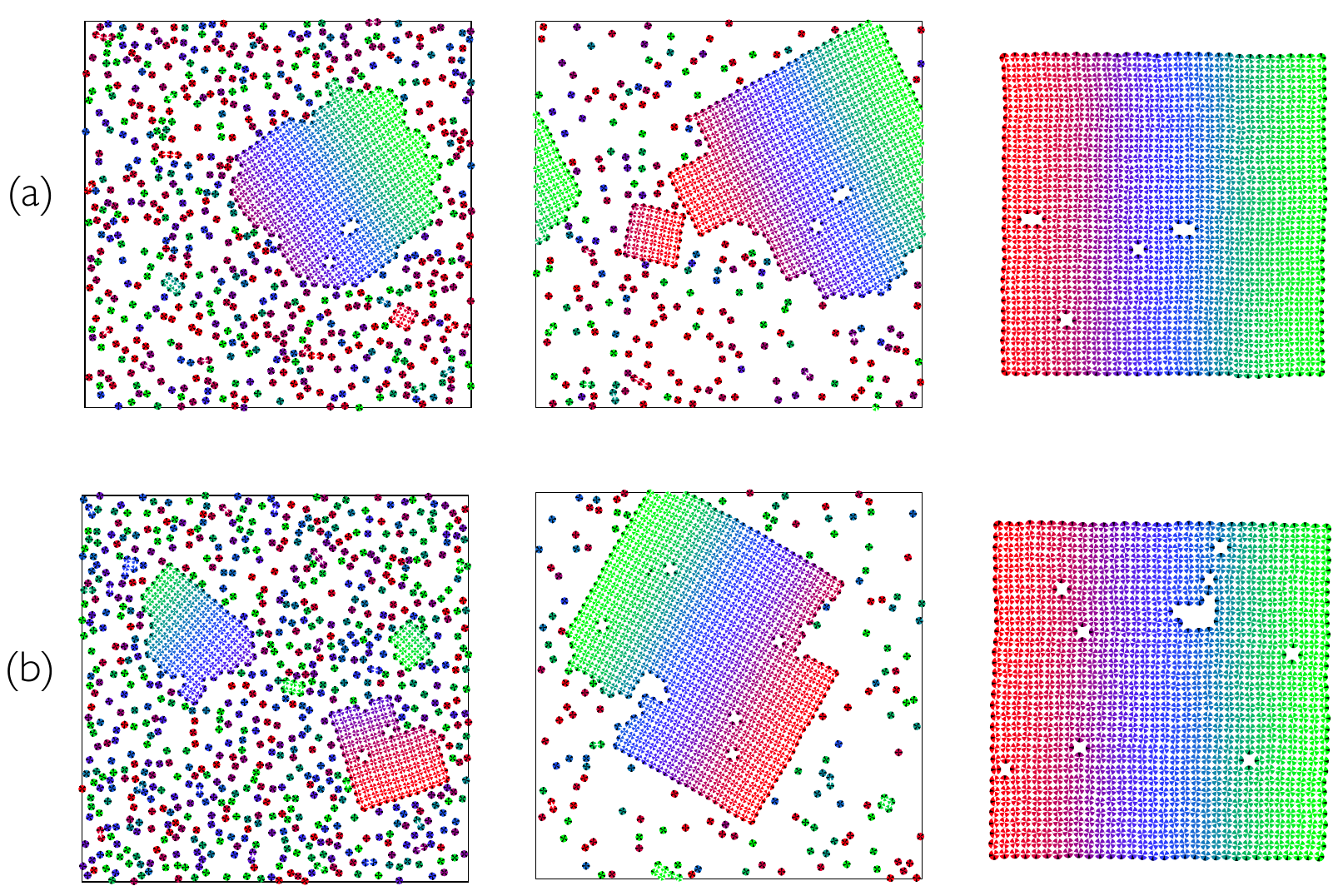}
\caption{\label{fig_non_hierarchical} Non-stage-by-stage self-assembly results in the target structure with only a few errors if reversible interactions are used. Here we show time-ordered snapshots from two simulations (one in (a), one in (b)) run with native interactions of energy $8.6 \, \kt$ (and with no non-native interactions). Some mistakes result when multiple nucleation events occur and the resulting large clusters fuse.}
\end{figure*}

{\bf Stage-by-stage assembly is vulnerable to kinetic traps associated with non-native attractions.} We have shown that the stage-by-stage assembly scheme works in the presence of strong native interactions. However, its major weakness is that it is susceptible to kinetic traps associated with non-native interactions. The red line \f{fig_hierarchical_graph_time}(b) corresponds to simulations done in the presence of non-native interactions of energy $\kt$, meaning that such an attraction exists between every patch-pair in the system that does not participate in a native interaction. Native contacts are again unbreakably strong. At stage 4 there is a slight slowing of assembly, relative to the case of zero non-native interaction, because of brief mutual adhesion of large clusters.

For stronger non-native interactions we found that assembly is arrested prior to stage 4. In \f{fig_undesigned_graph}(a) we show, as a function of time, the size $N_{\rm max}$ of the largest natively-bonded cluster in the system, from independent trajectories run in the presence of attractive non-native interaction energies of $0$, $2 \, \kt$, $3\,  \kt,$ and $5 \, \kt$. In all cases, native contacts are unbreakably strong. Assembly in the presence of non-native interactions becomes very slow at some point, the earlier the stronger the interaction. In panel (b) we show the fraction $f \leq1$ of possible native contacts that have been made at time $t$ for all trajectories; these fractions show behavior similar to that of $N_{\rm max}$.

The cause of this dynamical slowdown can be seen clearly in configuration snapshots. In \f{fig_undesigned}(a) we show a time-ordered series of snapshots from a simulation run in the presence of attractive non-native contacts of energy $3 \, \kt$. The square-upon-square hierarchical assembly process is subverted by a second organizational process, that of square 4-mers into a structure in which cluster types are arranged essentially at random (compare the color pattern to that of the target structure). As a result, 4-mers combine to form natively-bonded higher-order structures only slowly. In panel (b) we show pictures of structures assembled in the presence of attractive non-native contacts of energy $2 \, \kt$, $3 \, \kt$, and $5 \, \kt$. These pictures show that assembly of compositionally-random structures occurs from smaller building blocks as non-native contact energy decreases. Non-native interactions of energy $5 \, \kt$ cause a compositionally-nonspecific assembly of monomers, so preventing even stage 1 of the hierarchical assembly process from happening.

{\bf `Conventional' self-assembly does not work in the presence of arbitrarily strong native interactions.} The hierarchical assembly procedure's chief virtue is its ability to tolerate arbitrarily strong native interactions. This ability is not common in molecular self-assembly, and is not shared by a self-assembly process involving the same set of components whose native interactions are all turned on from the start of the simulation. A typical result of this process is shown in \f{fig_sequential_strong}. Even though only native interactions operate, if these form in the wrong order -- which, inevitably, they do -- then the target structure cannot assemble because of steric obstructions. 

However, conventional self-assembly done in the presence of reversible native interactions can result in the target structure with few errors. By running a large number of simulations with different values of the native interaction energy, we found that, for native interaction energy of 8.6 $\kt$, nucleation happened after some delay but was still rapid enough to be seen in direct simulation. Data from two such simulations are shown in \f{fig_non_hierarchical} and \f{fig_non_hierarchical_graph}. Some mistakes result when multiple nucleation events occur and the resulting large clusters fuse in a way that blocks certain binding sites. But assembly is largely successful, reiterating the results of Refs.~\cite{reinhardt2014numerical} and~\cite{ke2012three}: nucleation and growth of an `information-rich' structure of about $10^3$ components can occur in the presence of reversible interactions of fixed strength.

\begin{figure}[t]
\centering
\includegraphics[width=\linewidth]{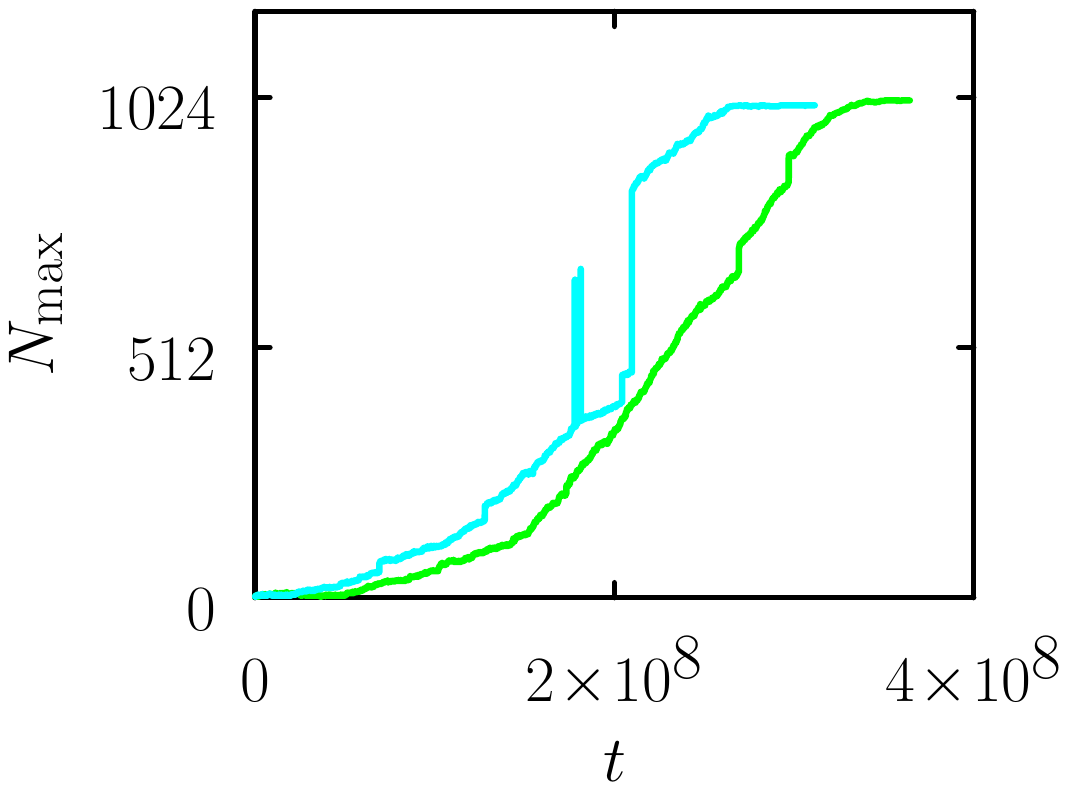}
\caption{\label{fig_non_hierarchical_graph} Size $N_{\rm max}$ of the largest natively-bound cluster in the system, for the two trajectories shown in \f{fig_non_hierarchical}. The green (resp. cyan) line here corresponds to the top (resp. bottom) panel in \f{fig_non_hierarchical}. The final yield of assembly is slightly lower for the case of the cyan/bottom trajectory.}
\end{figure}

\section{Conclusions}
\label{conc}

Ensuring the thermodynamic stability of a structure made from $Q$ distinguishable components requires the energy scale of inter-component interactions to grow in proportion to $\kt \ln Q$. But self-assembly done in the presence of strong interactions results in general in kinetic trapping, suggesting a limit to the size of a self-assembled structure built from distinguishable components. Here we have used simulation to show that the stage-by-stage assembly of squares demonstrated in experiment in \cc{park2006finite} can work even in the presence of arbitrarily strong interactions, as long as those interactions are the `native' ones required to stabilize the target structure. This property is not shared by most other forms of self-assembly, and is required if one wishes to make large, information-rich structures in equilibrium.

Note that we have assembled a single structure of size $2^{10}$ to demonstrate that error-free hierarchical assembly can occur even in the face of arbitrarily strong interactions. We stopped at stage 5 of the procedure because simulating larger systems is too time-consuming, but the hierarchical procedure can be used in principle to assemble a single ``information-rich'' structure of any size. One way to see this is to consider a visual proof by induction. \f{fig_hierarchical} shows the hierarchical assembly of a $2^{10}$-particle structure from monomer building blocks. Imagine now that we have $2^{10}$ such reactions occurring in parallel. Provided that disc-disc native interactions are chosen in the appropriate manner, combining the results of all reactions would lead to the hierarchical assembly of a $2^{20}$-particle structure: imagine a time-series like that shown in \f{fig_hierarchical} in which the smallest discs shown are not monomers but the $2^{10}$-particle assemblies that result from the first set of reactions. And so on. Size limitations then result from practical considerations.

One such consideration is that the stage-by-stage scheme is particularly susceptible to kinetic trapping caused by attractive non-native interactions, and so these must be suppressed if the scheme is to be used to make large objects. With DNA-mediated interactions alone, as noted in \cc{frenkel2015order}, it seems difficult to guarantee strong native interactions and very weak non-native ones, on account of accidental basepair complementarity. Choosing complementary sequences at random, one would expect two `non-complementary' 4-letter sequences of length $M$ to be `accidentally' complementary in $M/4$ places on average. To suppress non-native interactions it may be necessary to equip components with repulsive interactions intermediate in energy scale between fully-complementary interactions and accidentally-complementary ones.

If this can be done, then in principle the scheme presented in \cc{park2006finite} and simulated here can be used to make large objects. However, technical challenges exist, of which four are as follows (more detailed discussions of several of these points can be found in the DNA tile literature\c{demaine2011one,patitz2010self,abel2010shape,park2006finite}). First, one must be able to `switch on' interactions selectively. Perhaps if native interactions correspond to a certain number of fully-complementary basepairs, then selected strands could initially be `turned off' or `protected' by partially-complementary single strands. At the required stage of the process the system temperature could be raised slightly, so as to allow thermal dissociation of protector strands only. Protector strands of subsequent stages must be successively more strongly bound, and all must be much less strongly bound than native inter-component bonds. Second, one must be able to monitor the progress of assembly with sufficient precision to know when all the structures in the current stage have assembled (or wait long enough to be sure that this has happened); premature activation of the next stage's bonds could cause kinetic trapping. Perhaps `smarter' ways to vary inter-component interactions without requiring explicit intervention by the user can be developed, by using novel feedback schemes\c{klotsa2013controlling} or other nonequilibrium controls of assembly\c{byun2013unconventional,han2013large,li2014crafting}. Third, large clusters diffuse slowly, particular when bound to a surface, and so it is likely that stirring the solution or shaking the surface will be required to allow assembly to proceed beyond a certain lengthscale (shaking may also help break up kinetic traps caused by residual non-native interactions).  Fourth, this scheme confronts the challenges inherent to any scheme involving a large number of component types, such as difficulties of synthesis (the number of DNA labels must increase with system size, for instance; see the SI of \cc{park2006finite}) and issues of long timescales. That said, \cc{park2006finite} shows that square-upon-square assembly reactions can be run in parallel: one could in principle combine the outcome of $K$ separate hierarchical processes as the starting point for the next stage of assembly, thereby achieving $K$-fold speed-up relative to the case of sequential self-assembly. Table~\ref{table_summary} compares conventional and hierarchical assembly schemes.

We note finally that the scheme proposed here cannot be applied (without introducing the idea of twist-sensitive, protein-like interactions\c{mannige2009geometric,levy20063d}) to two-dimensional objects free to move in three-dimensional space: flip over a square and it can bind with its intended partner in a non-ideal way. A similar observation applies to three-dimensional structures. Therefore, we speculate that building 3D information-rich structures may be best done by using a hierarchical scheme to generate large 2D surface-bound assemblies, and then building layer by layer on top of them.

\begin{table*}[ht]
\caption{Pros and \con{contras} of assembly schemes}
\centering
\begin{tabular}{c |c}
\hline
conventional & stage-by-stage \\ 
\hline
\con{Intolerant of strong native interactions} & Tolerant of strong native interactions\\
Tolerant of weak non-native interactions & \con{Intolerant of weak non-native interactions} \\
No need for user intervention & \con{Requires user intervention} \\
\con{Reaction not easily parallelizable} & Reactions can be run in parallel \\
\hline
\end{tabular}
\label{table_summary}
\end{table*}

\acknowledgements

This work was done at the Molecular Foundry at Lawrence Berkeley National Laboratory, supported by the Office of Science, Office of Basic Energy Sciences, of the U.S. Department of Energy under Contract No. DE-AC02--05CH11231.

\appendix
\section{Note on the consequences of particle distinguishability}

Imagine that our goal is to promote the self-assembly, from solution, of a solid, equilibrium structure made of $Q$ components. Components generally lose translational and rotational entropy upon going from solution to a solid structure, and so the energy scale $\epsilon/(k_{\rm B}T)$ of inter-component bonds in the solid structure must be large enough that a monomer's energetic gain upon solidification exceeds temperature times the entropic cost of its removal from solution. If components are distinguishable, and if the solid structure of interest conforms to one particular arrangement of components, then upon solidification one {\em also} loses an entropy $Q^{-1} k_{\rm B} \ln Q! \approx k_{\rm B} (\ln Q-1)$ per particle, because the number of accessible configurations of the solution phase is $Q!$ times that of its distinguishable counterpart. The inter-component bond strength $\epsilon/(k_{\rm B}T)$ must therefore grow on the scale of $\ln Q$ in order to ensure the stability of a distinguishable structure built from $Q$ components. The average bond energy per particle must be at least $\kt/(z/2) \times \ln Q$ greater than in the corresponding single-component case, where $z$ is the number of contacts made by each particle. For $Q$ macroscopic, of order $10^{24}$, this excess bond energy exceeds $27 \,\kt$ when $z=4$ (appropriate to the case modeled here or the experiments of \cc{ke2012three}), and is about $9 \, \kt$ for $z=12$. 

A similar result can be obtained by considering the microscopic dynamics of growth. Say that in the indistinguishable case the rate of arrival of a component at a given point on a solid structure is $R_{\rm arrive}$, which is proportional to component concentration. Say that the rate of departure of that particle from the structure is $R_{\rm depart} \propto \exp(-\beta z \epsilon)$, where $\beta \equiv 1/(k_{\rm B}T)$ and $z$ is the number of bonds made by the particle. In general, the arrival rate must exceed the departure rate in order to ensure stability of the structure with respect to dissolution. Now in the {\em distinguishable} case the arrival rate, for fixed total component concentration, is $\approx R_{\rm arrive}/Q$, because only about 1 in $Q$ encounters involves the component designed to fit at that point on the structure (now we imagine for simplicity that the number of components in solution is very large, and does not diminish upon binding). So upon going from indistinguishable to distinguishable the rate of arrival of material has dropped by a factor of $Q$. To guarantee structural stability, therefore, the rate of departure of material must drop by a similar factor, i.e. $R_{\rm depart} \to R_{\rm depart}/Q$. Hence stability requires that $\beta z \epsilon$ must increase in proportion to $\ln Q$. 

We tested this expectation by building the 1024-particle target structure and simulating it (in the same square simulations boxes that we used for dynamic simulations) in the presence of native interactions only, and in the presence of native and nonnative interactions of equal strength. In the former case particles make energetic bonds only if they adopt one particular position in the structure, while in the latter case the energy of the system is invariant under particle permutation, or if particles are rotated by multiples of $\pi/2$. We  therefore expect the extra pair energy (in units of $\kt$) required to stabilize the structure in the native-interaction case to be $\Delta \epsilon = \frac{1}{2} Q^{-1} \ln \left(4^Q Q!\right) \approx \frac{1}{2} \left( \ln 4 + \ln Q -1 \right) \approx 3.66$ (for $Q=1024$). By performing simulations for different values of binding energies we determined that in the presence of native interactions the structure melted for bond energy values $\epsilon$ between 7.3 and 7.1, while in the presence of native and nonnative interactions of equal strength the condensed structure melted for values of $\epsilon$ between 3.8 and 3.6. Although simulations of this nature do not give us a precise measure of structural stability, the difference between these two `melting points' is $3.5 \pm 0.2$, consistent with our expectation of 3.66.

The same considerations also suggest that the growth rate of a distinguishable structure will be a factor of $Q$ less than that of its indistinguishable counterpart, i.e. $R_{\rm arrive}-R_{\rm depart} \to Q^{-1} (R_{\rm arrive}-R_{\rm depart})$ upon going from indistinguishable to distinguishable.\\

\section{Stage completion times}
\label{rates}

Let us assume that collisions between squares are uncorrelated events. During stage $n$ of assembly these events occur with mean time $\tau(n) \propto 1/D(n-1)$, where $D(n-1) \propto 2^{-n}$ for translation-limited events, and $D(n-1) \propto 2^{-3n}$ for rotation-limited events. To go from stage $n-1$ to stage $n$ of assembly there must occur of order $2^{10-2(n-1)} \propto 2^{-2n}$ events. So the mean time to complete stage $n$ of assembly scales, very roughly, as $2^{-n}$ if assembly is translation-limited, and  $2^n$ if assembly is rotation-limited. The data of \f{fig_hierarchical_graph_time}(b) lie between these extremes but appear to generally increase with $n$, suggesting that assembly is dominated, especially at late stages, by the rotational component of diffusion.


\end{document}